\documentclass[aps,pre,onecolumn,superscriptaddress,floatfix,nofootinbib,longbibliography]{revtex4-2}
\usepackage{graphicx}
\usepackage{amsmath,amssymb,bm}
\usepackage{booktabs}
\usepackage{dcolumn}
\usepackage{xcolor}
\usepackage{placeins}
\usepackage[colorlinks=true,linkcolor=blue,citecolor=blue,urlcolor=blue]{hyperref}
\newcommand{\Fr}{\mathrm{Fr}}
\newcommand{\OmH}{\Omega_H}
\newcommand{\rh}{r_h}

\newcommand{\rev}[1]{#1}
\newcommand{\revr}[1]{#1}
\begin{document}
\title{Shallow-water black holes and lattice superradiance}
\author{Antonino Flachi}
\affiliation{Department of Physics \& Research and Education Center for Natural Sciences, Keio University, Yokohama, Japan}
\date{}
\begin{abstract}
Rotating black holes can amplify waves that scatter off them, a process known as superradiance that fluid analogues bring into the laboratory. Separating the horizon's role from dissipation in the surrounding fluid remains difficult. Here we realise a rotating analogue black hole in a lattice fluid: a shallow-water vortex whose horizon is set by a topographic sill, so that the only sink lies behind a supercritical layer that lets nothing out. We measure superradiance as energy extraction: the flux at the horizon is outward for co-rotating waves whose pattern rotates more slowly than the horizon, and inward for faster waves, counter-rotating waves and a non-rotating hole. A scan over three lattice viscosities separates the extraction at the horizon, which is inviscid, from that of the viscous fluid inside the corotation radius: well below the band edge the horizon is the absorber; near the edge the viscous share is half at the operating viscosity and vanishes at the lowest. A drain-anchored horizon, as in water-tank experiments, gives the same signs. Our analysis gives insight into the mechanism of superradiance and offers a cheap, flexible platform for designing analogue-gravity experiments.
\end{abstract}
\maketitle

\section*{Introduction}
Zel'dovich showed that an \emph{absorbing} cylinder rotating at angular velocity
$\Omega$ amplifies an incident wave of frequency $\omega$ and azimuthal number
$m$ whenever $\omega<m\Omega$~\cite{Zeldovich1971,Zeldovich1972}. In the frame
co-rotating with the absorber, whenever the wave has frequency
$\omega'=\omega-m\Omega<0$, it carries negative energy, so absorbing it
\emph{lowers} the energy of the field at the cylinder and the reflected wave
leaves with a larger amplitude than the incident one. For a black hole rotating with (horizon) angular velocity $\OmH$, the absorber is the horizon and the condition becomes
$\omega<m\OmH$~\cite{Misner1972,Starobinsky1973,Bekenstein1998,Brito2015}; the same mechanism
underlies the Penrose process~\cite{Penrose1969,Penrose1971} and, once the amplified wave is confined by a
mirror or a mass term, the black-hole bomb~\cite{Press1972,Cardoso2004bomb}.

Laboratory observations remain few. Torres \emph{et al.} measured the reflection
of
surface waves from a draining vortex in a water tank, finding
$|R|>1$ for co-rotating modes and $|R|<1$ for counter-rotating ones, with a
maximum amplification of $(14\pm8)\%$ at $3.7$~Hz~\cite{Torres2017}; this was the
first observation of rotational superradiance. Cromb \emph{et al.} sent
acoustic beams carrying orbital angular momentum $\ell$ through a rotating
absorbing disk and measured a transmission gain of up to $30\%$ once
$\omega<\ell\Omega$~\cite{Cromb2020}, the first realisation with a 
rotating absorber rather than a rotating flow. Braidotti \emph{et al.} focused a
weak beam carrying orbital angular momentum onto the core of a pump vortex in a
photon superfluid and measured over-reflection: negative-norm modes are generated
and trapped in the vortex core by a four-wave mixing process whose phase relation
is the Zel'dovich--Misner condition~\cite{Braidotti2022}. 
The electromagnetic setting that Zel'dovich had in mind was reached last, through induction machines~\cite{Braidotti2024,Cromb2025}, where the black-hole bomb was also realised~\cite{Cromb2025}; its nonlinear end state has since been observed on a draining vortex in superfluid helium~\cite{Svancara2026}. A Floquet analogue, in which the rotation is synthesised by a travelling-wave modulation and the gain is parametric, has been reported in a ring of time-modulated resonators~\cite{Nasari2026}.

In the experiments~\cite{Torres2017,Cromb2020,Braidotti2022,Braidotti2024,Cromb2025,Svancara2026} the absorption takes place in a
rotating frame: the absorber and the medium in which the wave is dissipated turn
together at some $\Omega_{\rm abs}$; the role of the boundary condition at the absorber has been analysed~\cite{Richartz2009}, and the conditions for detecting rotational superradiance in a fluid have been set out~\cite{Cardoso2016}.
For a solid absorber the identification is immediate: the foam disk or the metal
cylinder spin at $\Omega$, the wave is dissipated inside it, and the relevant
frame is the body's rest frame. For a drain or a vortex core the absorber is the fluid itself: the wave is removed by being carried across the horizon, and the fluid there rotates at the horizon angular velocity $\OmH$, so that
$\Omega_{\rm abs}=\OmH$. In the Floquet analogue~\cite{Nasari2026} the role of $\Omega_{\rm abs}$ is played by the speed of the modulation. In every case $\Omega_{\rm abs}$ is fixed by the apparatus, and gain appears when the superradiance condition is met. 
In a vortex flow, the horizon and the dissipation share the same region, the vortex core, so their contributions to the absorption are not separated. The only attempt we are aware of to realise and measure superradiance on a lattice is a gas-like model~\cite{Flachi2026}, where the measurement was inconclusive.

On a lattice, $\Omega_{\rm abs}$ is not known in advance. The drain is neither a body nor a fluid but a rule acting on the populations that carry both the wave and the flow, and the horizon spans only a few cells, where the lattice fluid need not behave like the continuum one. Whether a lattice hole superradiates must therefore be measured. What the measurement can show depends on where the dissipation sits relative to the horizon. In the inviscid problem, the exterior flux is fixed by a single condition at the horizon: the wave is purely ingoing there. Inside, the flow is supercritical and nothing returns, so dissipation confined behind the horizon is invisible from outside and the horizon is the absorber. Dissipation located outside the horizon is different: it acts where the exterior flux is formed, and the flux cannot separate it from the extraction.

We expect a lattice model of a drain to be of the second kind: its removal region extends across the horizon, as the vortex core does in the tank. A measurement there reads the horizon and the drain together; it can show that energy is extracted, but not what extracts it. A topographic sill removes this ambiguity by placing the sink behind the horizon. Where the bed rises, the layer thins and the inflow overtakes the waves: the horizon is a geometric crossing with no source term, and the only sink lies behind a supercritical layer. Bulk viscosity remains, and inside the corotation radius, where the wave carries negative energy, it is itself a rotating absorber; its share is separated by varying the viscosity. Flume experiments form their horizons similarly, over an obstacle in a channel~\cite{Weinfurtner2011,Euve2016,Euve2020}. Here a sill is used to separate the horizon of a rotating hole from its dissipation.

We read superradiance in the sense of Zel'dovich's argument, as negative absorption: an outward energy flux at the absorber for $\omega<m\Omega_{\rm abs}$, measured here on rings next to the horizon. The result is, to our knowledge, the first lattice construction in which rotational superradiance is realised and directly measured.

\section*{Results}

\subsection*{A shallow-water fluid on a lattice}

In the shallow-water limit, where the wavelength is much larger than the depth $h$, surface waves propagate at the local speed $c=\sqrt{gh}$ and obey $\omega=\mathbf v\cdot\mathbf k\pm c|\mathbf k|$, the dispersion relation of a massless field on the acoustic metric of the flow~\cite{Unruh1981,Schutzhold2002,Fischer2002}. 
In a draining vortex the horizon lies where the radial inflow reaches the wave speed, $|v_r|=c$, at radius $\rh$, and it rotates at $\OmH=v_\theta(\rh)/\rh$. The ergosurface lies where the total speed reaches the wave speed, $|\mathbf v|=c$, at a larger radius. Between the two a wave can escape but cannot remain at rest or counter-rotate: this is the analogue of the Kerr ergoregion. In the tank experiments~\cite{Torres2017,Torres2020} the absorber is the vortex core.

We use a lattice-Boltzmann model on the D2Q25 square lattice. Each cell holds populations $f_i$ that stream to neighbouring cells along fixed velocities $\mathbf e_i$ and relax towards a local equilibrium.

A force term makes this gas behave as a water layer. A lattice gas at rest has the pressure $c_s^2h$ of an isothermal gas, whereas a shallow layer of water has the hydrostatic pressure $gh^2/2$ and is pushed downhill where the bed slopes. We therefore add the body force $\mathbf F=(c_s^2-gh)\nabla h-gh\nabla b$. Its first term replaces the gas pressure gradient by the hydrostatic one; its second is the weight of the layer projected on the slope of the fixed bed profile $b(\mathbf x)$.

\begin{figure}[tbp]
\centering
\includegraphics[width=0.98\textwidth]{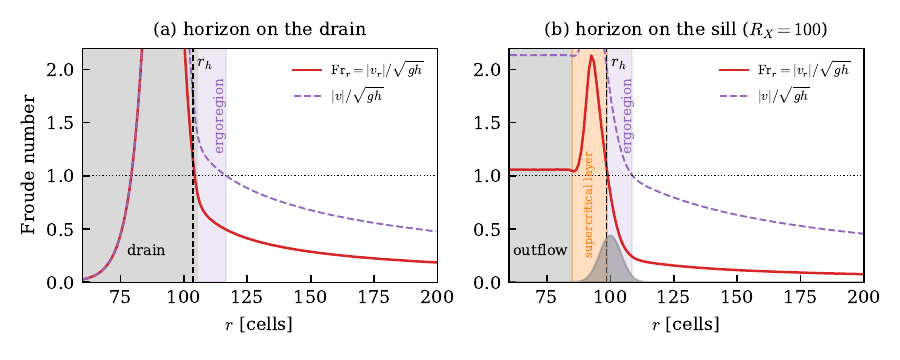}
\caption{\label{fig:matrix}Measured radial and total Froude numbers of the settled shallow-water flows ($B=30$). (a) Horizon on the drain ($A=40$), with the layer thinning into the drain (grey). (b) Horizon on the sill at $R_X=100$: the sill (dark grey, right axis) chokes the flow at its crest, where $\Fr_r=1$; inside it lie the supercritical layer (orange) and the free overfall (grey). The horizon $\rh$ and the ergoregion (purple) are marked.}
\end{figure}

On the lattice the surface-wave speed is emergent and local, as in the tank, so a horizon can form only where the flow thins the layer. 
Inside a horizon the radial inflow exceeds the wave speed, and with the swirl the total Froude number $\mathrm{Fr}=|\mathbf v|/\sqrt{gh}$, the Mach number of the surface waves, rises well above one (Fig.~\ref{fig:matrix}). Most lattice-Boltzmann shallow-water schemes run well below this, so neither the emergence of a horizon nor its stability is guaranteed. We chose this model because it remains linearly stable up to $|\mathbf v|/c_s\simeq1.3$, while the flows used here reach at most $|\mathbf v|/c_s\simeq1$. The parameters and the behaviour of the scheme at the crossing are given in Appendix~\ref{app:methods}. Throughout, lengths are in lattice cells and times in steps. 

The lattice fluid is not a discretisation of the shallow-water equations. It is defined by its microscopic rules, and it reduces to the continuum equations only where the flow is smooth on the cell scale and slow compared with $c_s$. A black-hole flow meets these conditions in the exterior, but not at the horizon crossing or inside it, so what the continuum predicts on a given background need not occur on the lattice. This distinguishes the present work from simulations that evolve a wave equation on a prescribed flow~\cite{Cherubini2005,Federici2006}.

\subsection*{Horizons, with and without sills}
Our model realises a horizon in two ways (Fig.~\ref{fig:matrix}). In the first, the horizon forms on a drain, as in the tank. In the second, a sill on the bed detaches the horizon from the exit. Both are measured in the same way. The drain reproduces the configuration of the tank; the sill isolates the horizon from the sink, and is where we establish the origin of superradiance.

On the drain, a reservoir supplies a nominal volume flux $2\pi A$ with circulation $2\pi B$, and water is removed at every step inside a central disc of radius $r_d=100$ (in lattice units). Between the two the flow evolves freely: it converges, thins the layer and reaches the wave speed at $\rh\simeq104$, on the edge of the drain. The removal therefore acts on the horizon cells and on the subcritical cells just outside them. Absorber and dissipation coincide, as they do at the vortex core of the tank, and a flux measured there reads both at once, with no way to separate them.

The sill moves the horizon away from the exit. The drain is replaced by a smooth circular bump on the bed, a sill with its crest at radius $R_X$, and the water leaves over a free overfall at a brink inside the crest. If the sill is high enough, the flow accelerates over it and reaches the wave speed at the crest. The crest chokes the flow and selects the flux, and the crossing is smooth: subcritical outside the crest, supercritical inside. In this supercritical layer waves can only move inward, so the exit is hidden behind the horizon. The circulation and the exit are those of the tank; only the horizon has moved.

This construction separates the horizon from every sink. The sill itself is conservative, so the horizon is a purely geometric crossing on which no source term acts. The only sink is the outflow at the brink, which lies behind a supercritical layer ($14$ cells wide in our simulation). This layer is effectively one-way for the long waves used here: a depth perturbation introduced inside it delivers to the region outside the horizon at most $10^{-4}$ of the energy delivered by the same perturbation introduced outside. The flux measured outside the horizon is therefore independent of the conditions at the exit. What remains outside is the bulk viscosity of the lattice fluid, which inside the corotation radius $r_c$, where $\omega<m\,v_\theta/r$, absorbs a wave of negative energy and so extracts as the horizon does; its share is isolated below by varying the viscosity.

We use two sills, $R_X=100$ and $115$, which at the same circulation give the horizon angular velocities $\OmH=3.0\times10^{-3}$ and $2.3\times10^{-3}$. Every rotating flow has a non-rotating twin ($B=0$) as reference. Outside the ergoregion the sill and drain flows have the same swirl and wave speed; the drain carries a larger radial inflow. The essential difference lies behind the horizon. The main measurements are made on the sills; the drain, measured in the same way, provides the comparison with the tank. Parameters, the exit condition and the comparison of the two flows are given in Appendices~\ref{app:methods}, \revr{\ref{sm:flow} and \ref{sm:horizon}--\ref{sm:gain}}.

\subsection*{Measuring superradiance on a lattice}
Superradiance is energy extraction. For $\omega<m\Omega_{\rm abs}$ the absorber takes in a wave of negative energy, the field outside it gains, and the gain appears as an outward energy flux at the absorber. In a lossless exterior the extracted energy returns as an amplified reflected wave, and the usual measure is the reflection coefficient $R$: the ratio $|R|^2$ of reflected to incident energy exceeds one.

On the lattice $|R|^2$ is out of practical reach, for a physical reason. A pattern with $m$ crests around a circle of radius $r$ travels as a wave only if its frequency in the frame of the swirling fluid exceeds $mc/r$,
\begin{equation}
\omega-\frac{m v_\theta}{r}>\frac{mc}{r}.
\label{eq:travel}
\end{equation}
Superradiance requires $\omega<m\OmH$. The two conditions are compatible only where $(v_\theta+c)/r<\OmH$, which fails at the horizon, where $v_\theta/r=\OmH$, and holds only far from the hole. A wave in the superradiant band, $\omega=fm\OmH$ with $f<1$, therefore travels only outside the turning radius $r_t\simeq(v_\theta+c)/(f\OmH)$ and is evanescent inside it. Equation~(\ref{eq:travel}) neglects the radial inflow; with it the condition reads $(\omega-mv_\theta/r)^2>(m^2/r^2)(c^2-v_r^2)$, which always holds at the horizon, so the wave travels again in a thin layer next to it (about two cells at $f=0.5$, $m=4$ on $R_X=115$) and tunnels through the barrier in between.

A reflection measurement thus needs a box large enough to contain the travelling region, with several radial wavelengths between the drive and the turning point. Even in such a box, the horizon's in-band contribution to the far-field reflection is more than an order of magnitude smaller than the viscous attenuation of the wave across the travelling region, and $|R|^2$ cannot be extracted. We therefore measure the extraction where it is produced, next to the horizon. The wave is driven from a thin ring at $r=170$, outside the ergoregion but inside the turning radius at every in-band frequency used on the sills.

On the drive ring the depth is modulated at every step by a small fraction $\epsilon\cos(m\theta-\omega t)$, a pattern of $m$ crests rotating at angular velocity $\omega/m$ in the sense of the vortex. The modulation is kept on until the response is steady. We then record, on rings just outside the horizon, the energy crossing each ring per unit time, $I(r)=2\pi r\langle(g\,\delta h+\mathbf v_0\!\cdot\!\delta\mathbf u)(h_0\delta u_r+\delta h\,v_{0r})\rangle$, where $\delta h$ and $\delta\mathbf u$ are the departures from the settled flow $(h_0,\mathbf v_0)$ and outward flux is positive (by convention). The net flux $I_{\rm net}$ is the mean of $I$ over these rings. $I_{\rm net}<0$ means that the hole absorbs; $I_{\rm net}>0$ means that it gives out more energy than it receives, which is superradiance. The flux of angular momentum $F_J(r)$ satisfies the single-mode relation $F_J=(m/\omega)\,I$ to within $7\%$ for $m=4$ at every frequency away from the band edge on both sills (Table~\ref{tab:fj}), which confirms that the measured flux is carried by the driven mode.

\begin{figure}[tbp]
\centering
\includegraphics[width=0.98\textwidth]{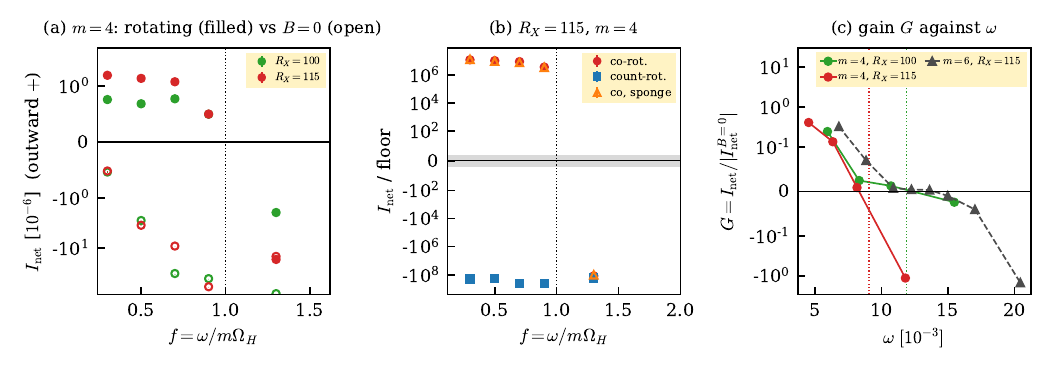}
\caption{\label{fig:ledger}Net flux $I_{\rm net}$ on the sills. (a) Co-rotating $m=4$ wave against $f=\omega/m\OmH$ on the two sills (filled) and their non-rotating twins at the same absolute frequencies (open). (b) $R_X=115$, $m=4$: net flux of the co-rotating (circles) and counter-rotating (squares) waves, in units of the undriven floor; triangles repeat the co-rotating measurement with the absorbing layer (sponge) at the outer boundary. (c) Near-field gain $G=I_{\rm net}/|I^{B=0}_{\rm net}|$ at equal drive against the absolute frequency $\omega$ (\revr{dotted lines: $4\OmH$ of each sill}). All runs without the sponge.}
\end{figure}

Each measurement has two controls. In the counter-rotating run the pattern turns against the vortex, and the wave must be absorbed. In the undriven run the modulation is off, and the flux carried by the settled flow defines the floor. A measurement is accepted only if $|I_{\rm net}|$ exceeds three times this floor; away from the band edge the fluxes exceed it by several orders of magnitude (Fig.~\ref{fig:ledger}b).

The outer boundary also requires care. The reservoir reflects outgoing waves, so some runs use an absorbing layer (sponge) at the outer edge, and its presence changes the magnitude of the fluxes. The boundary treatment cannot, however, produce a positive $I_{\rm net}$. The probe rings lie next to the horizon, well inside the drive, so anything arriving from outside, reflections included, crosses them inward and counts as negative. Only a wave of negative energy moving inward counts as positive, and this is the horizon's own signal: the transmitted wave, which on the sills can be followed through the horizon down to the brink. In the steady state the energy stored between the rings is constant, so the measured flux is energy passing through the rings, not energy accumulating between them. Farther out, between the horizon and the drive, the flux varies with radius, \revr{but this variation comes from dissipation in the fluid and from the boundary treatment, not from the hole.}

The flux does not give $|R|^2$, because incident and returned fluxes are not separated where nothing travels. We therefore compare the rotating hole with its non-rotating twin, driven identically, which can only absorb. The ratio $G=I_{\rm net}/|I^{B=0}_{\rm net}|$ is the near-field gain: the energy returned by the rotating horizon in units of the energy its twin absorbs for the same drive. $G>0$ signals superradiant extraction. 

We thus measure $G$ in place of $|R|^2$. In a lossless exterior the two are equivalent: the energy extracted at the horizon returns as an amplified wave, so $G>0$ means $|R|^2>1$. In a viscous fluid they are not, because the wave loses energy on its way out. We therefore measure the extraction next to the horizon and quantify the viscous losses separately; these show that the extracted energy is dissipated before it can form a travelling wave. 
The controls, the boundary treatment and the energy budget are detailed in Appendices~\ref{sm:ledger} and \ref{sm:gain}.

\begin{figure}[tbp]
\centering
\includegraphics[width=0.7\textwidth]{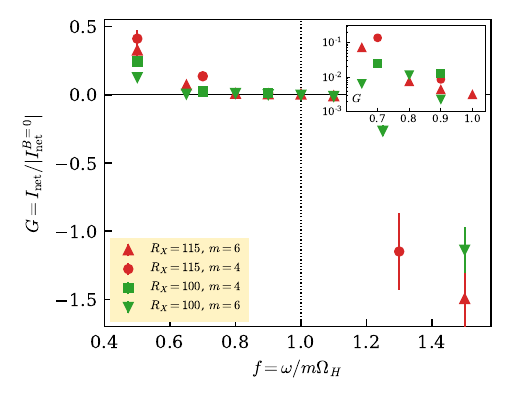}
\caption{\label{fig:gain}Near-field gain $G=I_{\rm net}/|I^{B=0}_{\rm net}|$ on the sills, from runs without the sponge (bars: window spread combined with a $15\%$ systematic for the depth mismatch between the rotating flow and its twin next to the horizon, or half the sponge-on/off difference where larger, Appendix~\ref{sm:gain}). Inset: $G$ for $0.6<f\le1$ on a logarithmic scale, positive values only; the negative values at $f=1.0$ on the $R_X=100$ sill lie outside it.}
\end{figure}

\subsection*{Superradiant energy extraction}
We measure the energy flux on the two sills for co-rotating waves with $m=4$ and $m=6$ (Fig.~\ref{fig:ledger}, Table~\ref{tab:flux}). The result is the same in every case. The net flux is outward below the band edge, $f=\omega/m\OmH=1$, and inward above it. The counter-rotating wave and the non-rotating twin give an inward flux at every frequency measured. All fluxes scale with the square of the drive, as expected of a linear response.

The band edge follows the horizon. It scales with $m$ and with $\OmH$, whose value differs between the two sills (Table~\ref{tab:flux}), and in all four cases the flux changes sign within $10\%$ of $m\OmH$ (the $m=4$ edge scans under the control protocol, which changes no sign). The angular velocity of the brink is \revr{about one third higher} than $\OmH$ and lies outside this interval. The band is therefore set by the horizon. The viscous share of the extraction does not move it: the annulus $\rh<r<r_c$ in which the viscous fluid extracts closes as $f\to1$, since $r_c=\rh/\sqrt f$, so the viscous and the horizon contributions vanish at the same frequency.

\begin{table}[t]
\caption{\label{tab:flux}Net energy flux next to the horizon on the two sills. Magnitudes are in units of $10^{-6}$, for a drive amplitude $\epsilon=2\times10^{-4}$. 
{``In band'' gives the range of the outward flux over the measured frequencies with $f=\omega/m\OmH<1$ ($f=0.3$ to $0.9$ for $m=4$), and ``above the edge'' the inward flux at the first measured frequency beyond it. The counter-rotating wave and the non-rotating hole are driven at the same absolute frequencies as the co-rotating wave; their columns give the range of the inward flux from $f=0.5$ upwards.
The last column is the near-field gain $G=I_{\rm net}/|I_{\rm net}^{B=0}|$ at $f=0.5$, \revr{the lowest frequency at which the twin's flux is reproduced to within $20\%$ under both boundary treatments.}} All rows are measured without the absorbing layer at the outer boundary (sponge); the $m=4$ runs repeated with it are in Table~\ref{tab:m4}. All frequencies are listed in \revr{Tables~\ref{tab:m4}--\ref{tab:m6x100}}, and the drain values in Table~\ref{tab:sink}.}

\centering\footnotesize\setlength{\tabcolsep}{3.5pt}
\begin{tabular}{llcccccc}
\hline
horizon & $\OmH\,[10^{-3}]$ & $m$ & co-rotating, in band & above the edge & counter-rot. & non-rot. & $G$ \\
 & & & (outward) & (inward) & (inward) & (inward) & at $f=0.5$ \\
\hline
sill $R_X=100$ & $2.98$ & 4 & $0.50$--$0.78$ ($f=0.3$--$0.9$) & $1.9$ ($f=1.3$) & $11$--$66$ & $2.8$--$80$ & $0.24$ \\
 & & 6 & $0.09$--$0.18$ ($f\le0.9$) & \rev{$0.03$ ($f=1.0$)} & \revr{$13$--$54$} & $0.7$--$76$ & $0.13$ \\
sill $R_X=115$ & $2.27$ & 4 & $0.50$--$1.7$ ($f=0.3$--$0.9$) & $17$ ($f=1.3$) & $23$--$55$ & $3.5$--$58$ & $0.41$ \\
 & & 6 & $0.18$--$0.34$ ($f\le0.9$) & $0.55$ ($f=1.1$) & \revr{$11$--$52$} & $0.9$--$75$ & $0.33$ \\
\hline
\end{tabular}
\end{table}

Figure~\ref{fig:gain} shows the near-field gain on the sills. In band $G$ is positive. It is largest at $f=0.5$, between $0.13$ and $0.41$, and decreases towards the band edge, where it vanishes and changes sign. Above the edge $G$ is negative: the rotating hole absorbs, \revr{increasingly with frequency, and at $f=1.5$ as much as its twin.}

As a check, we compared the measured fluxes with a continuum calculation: the linearised shallow-water equations, solved on the measured background with the regular condition at the horizon. Per unit depth amplitude on the drive side, the lattice flux next to the horizon lies within $20\%$ of the continuum value at every frequency, \revr{and the radial profiles of the amplitude agree to within $5\%$ at $f=0.5$ and $1.3$ and $14\%$ at $f=0.7$ between the read-out rings and $r=160$; at $f=0.9$ the lattice amplitude grows faster towards the drive, by up to $50\%$ at the highest viscosity.} \revr{These differences are expected, since} the lattice fluid is viscous and is not a discretisation of the continuum equations.

To separate the horizon from the viscous fluid, we repeated the measurement at three viscosities spanning a factor of four. The criterion is that the extraction at the horizon is inviscid and does not scale with $\nu$, whereas the extraction by viscosity inside the corotation radius does. In each case we compared the flux next to the horizon, normalised by the local depth amplitude, with the inviscid solution regular at the horizon. At $f=0.5$ the measured flux remains close to the inviscid value, and its largest departure, about one third at the lowest viscosity, appears equally in the absorbing twin (ratios $0.98/1.04$, $0.91/0.87$ and $0.67/0.68$ for the co-rotating wave and the twin) and is therefore unrelated to the extraction. At $f=0.9$ the co-rotating flux is $2.05$, $1.57$ and $1.00$ times the inviscid value as $\nu$ decreases, while the twin's stays at $1.0$: near the band edge half of the measured extraction is viscous at the operating viscosity, and none at the lowest. The two frequencies differ in where the corotation radius $r_c=\rh/\sqrt f$ lies: at $f=0.9$ it falls on the read-out rings ($r_c=121$), so the viscous layer sits on the probe, whereas at $f=0.5$ it lies forty cells away ($r_c=162$), eight cells inside the drive at $r=170$. {There the perturbation is $80$--$90\%$ vortical, and the flux changes there by exchanging energy with the flow's vorticity (Appendix~\ref{sm:ledger});} moving the drive to $r=140$ or $200$ leaves the signs unchanged. The gain keeps its sign at all three viscosities. Together, these results identify the horizon as the absorber well below the band edge, with the viscous contribution near the edge isolated by the scan.

We repeated the measurements on the drain, the configuration of the tank. The fluxes have the same signs as on the sills: outward in band and inward above the edge for the co-rotating wave, and inward for the counter-rotating wave and for the twin. The gain is of comparable size. When the drain is made smaller, $\OmH$ increases and the band edge follows. \revr{At $f=0.5$ and above the edge these signs are robust to the boundary treatment, the viscosity and the resolution; at $f=0.9$ the small in-band flux on the drain reverses at lower viscosity, where the removal region outweighs the extraction.} The comparison with the inviscid solution shows that, on the drain, the removal region absorbs together with the horizon, as expected from its position across it. The drain therefore confirms superradiant extraction in the configuration of the tank, while the sill identifies its origin.

The tank experiments use $m=1$ and $2$. Our main runs use $m=4$ and $6$, because lower $m$ require a larger simulation box and are computationally more costly. Our simulations of $m=2$, $3$ and $5$ give the same signs. The full frequency scans, the drain runs and the tests of viscosity, resolution and boundary treatment are given in Appendices~\ref{sm:ledger}--\ref{sm:bench}.

\section*{Discussion}

We have presented a lattice model in which rotational superradiance emerges dynamically from a shallow-water black hole. The flow, the horizon and the waves evolve together under the microscopic rules of the lattice, and the energy extracted at the horizon is measured directly in this evolving system. 

The model also provides a framework for analogue gravity. Bed profiles, horizons and boundary conditions can be changed at little cost, so that different realisations of shallow-water black holes, and of analogue systems more generally, can be designed and tested numerically before they are built. The sill introduced here is an example. In a tank, where horizon and dissipation coincide at the vortex core, a submerged sill with a supercritical layer between it and the drain would separate them in the same way. Once a configuration is fixed, its dissipation can be studied within the same framework. The model can also be realised as a tabletop experiment, as a synchronous digital circuit: a digital version of an analogue black hole, whose properties are probed electrically~\cite{Flachi2026b}. There a single flipped bit, which excites waves of every length, leaks through the supercritical layer by an amount that falls exponentially with depth: the one-way property of the layer holds for long waves.

Finally, the analysis gives insight into the mechanism of superradiance. By separating the horizon from the sink, and the horizon's extraction from that of the viscous fluid inside the corotation radius, the sill shows that energy is extracted at the horizon, in the frequency band set by Zel'dovich's condition, and that this extraction can be measured independently of the amplification of a travelling wave.

\begin{acknowledgments}
This work was conducted using the Miyabi Supercomputing System at the Joint Center for Advanced High Performance Computing (JCAHPC).\\
\emph{Data availability and AI disclosure.} All data used in this paper are available in the Zenodo repository \href{https://doi.org/10.5281/zenodo.23384373}{10.5281/zenodo.23384373}. AI tools were used to assist with code for the simulation sweeps, data analysis and figure plotting, and with language editing; all figures are plotted directly from the data in the repository. 
\end{acknowledgments}

\appendix
\section{Methods}\label{app:methods}
\subsection*{Lattice scheme}
The model is a lattice-Boltzmann scheme on the D2Q25 square lattice. Lattice spacing and time step are both unity, so that throughout the paper lengths are in cells, times in steps, and velocities, depths and bed heights in the corresponding lattice units.

The populations $f_i$ stream to neighbouring cells along the fixed velocities $\mathbf e_i=(e_x,e_y)$, with $e_x,e_y\in\{0,\pm1,\pm2\}$, and the lattice speed of sound is $c_s=1/\sqrt2$. The depth is $h=\sum_if_i$ and the momentum is $h\mathbf u=\sum_i\mathbf e_if_i+\tfrac12\mathbf F$, where $\mathbf F$ is the body force and the second term is the half-step correction of the forcing scheme~\cite{Guo2002}.

The equilibrium is the discrete Maxwell--Boltzmann distribution of the lattice, written as a product of one-dimensional expansions in Hermite polynomials of the velocity and truncated at fourth order. The same polynomials define the moments of the populations (depth, momentum, stress and higher), and the collision relaxes each moment at its own rate: $1/\tau$ for the second-order moments, which sets the viscosity $\nu=c_s^2(\tau-\tfrac12)$, and $s_h$ for all higher moments, which affects only stability. The lattice construction, the Hermite-expanded equilibrium and the moment-space collision follow the standard construction~\cite{Shan2010,ShanYuanChen2006,Coreixas2017}; earlier shallow-water lattice-Boltzmann schemes exist as well~\cite{Salmon1999,Zhou2004}.

The body force is $\mathbf F=(c_s^2-gh)\nabla h-gh\nabla b$. We use $\tau=0.70$, that is $\nu=0.10$, and $s_h=0.6$; the viscosity scan also uses $\nu=0.05$ and $0.025$. Further details of the model and of the numerical analysis are given in Appendices~\ref{sm:scheme}--\ref{sm:bench}.

\subsection*{Stability at the horizon crossing}
The horizon crossing is where most shallow-water lattice schemes fail, because the depth and the velocity change rapidly there. Along the converging flow the Bernoulli energy $E=|\mathbf v|^2/2+gh$ and the flux $q=rhv_r$ are conserved, so that on a flat bed, at each radius,
\begin{equation}
E(h)=\frac{q^2}{2r^2h^2}+gh+\frac{v_\theta^2}{2}.
\label{eq:bernoulli}
\end{equation}
The swirl term depends on $r$ only, so $E(h)$ has a minimum at the critical depth $h_c=(q^2/gr^2)^{1/3}$, where $v_r=c$. The depth profile $h(r)$ follows by inverting Eq.~\eqref{eq:bernoulli} at fixed $E$. {Near the minimum $E-E_{\min}\propto(h-h_c)^2$, so that $h-h_c\propto\pm\sqrt{E-E_{\min}}$: the critical depth is a square-root branch point. On a flat bed $E_{\min}(r)$ increases inward, so at fixed $E$ the subcritical and supercritical branches meet at a fold, where $dh/dr$ diverges, and the inviscid flow cannot continue; the lattice flow crosses only because the drain removes water there, and the depth falls within a cell of $\rh$, the velocity $v_r=q/(rh)$ rising with it. On the sill the bed adds $gb(r)$ to $E$, the minimum $E_{\min}(r)$ is stationary where $g\,b'=(v_r^2+v_\theta^2)/r$, and the crossing is regular there, on the inner flank of the sill: $0.6$ cells inside the crest in the inviscid estimate, and $0.7$--$1.0$ cells on the lattice (Appendix~\ref{sm:horizon}).}

The linear stability of the collision rule is a property of the lattice gas at its fixed $c_s$, so the relevant quantity at the crossing is the lattice Mach number $|\mathbf v|/c_s$, whatever the Froude number. The D2Q25 collision with its fourth-order equilibrium is linearly stable up to $|\mathbf v|/c_s\simeq1.3$ (Appendix~\ref{sm:scheme}). On the flows of this paper the wave speed at the crossing is $\sqrt{gh}/c_s\simeq0.25$ on the sills and $0.3$ on the drain, and $|\mathbf v|/c_s$ reaches $0.6$ inside the horizon on the sills and $0.8$--$1.0$ on the drain. The flow therefore passes through the crossing without special treatment; multispeed lattices reach transcritical shallow-water flow for the same reason~\cite{LaRocca2015}. The margin to the stability limit is what bounds the depth contrast and the swirl that can be imposed.

\subsection*{Simulation box, reservoir and sponge}
The simulations use a square of $768^2$ cells centred on the hole. The disc $r<r_{\rm out}=300$ evolves freely. The cells outside it, $r>r_{\rm out}$, form a reservoir: at every step their populations are overwritten with the equilibrium $f_i^{\rm eq}(h_{\rm res},\mathbf v_{\rm res})$ of a fixed state of depth $h_{\rm res}=2$ and velocity $\mathbf v_{\rm res}=-(A/rh_{\rm res})\hat{\mathbf r}+(B/r)\hat{\bm\theta}$. Populations streaming from the reservoir into the disc carry the reservoir's depth, a nominal volume flux $2\pi A$ and a circulation $2\pi B$; populations streaming out of the disc are discarded. Inside the disc the flow selects itself, thinning the layer as it converges. We set $g=0.05$ and $B=30$, with $A=20$ on the sills and $A=40$ on the drain; the non-rotating twins have $B=0$.

The reservoir reflects outgoing waves. The measurements on the sills are made with no treatment of the outer boundary; as a control, every $m=4$ run was repeated with a damping layer (sponge) in $250<r<300$, which acts on the wave part of the velocity only and leaves the steady flow and the horizon unchanged (Fig.~\ref{fig:ledger}b; Appendix~\ref{sm:sponge}). The measurements on the drain use the sponge.

\subsection*{Flow construction and measurement}
On the drain, water is removed at every step by the rule $f_i\to(1-s\,\chi(r))f_i$. The profile $\chi$ equals $1$ inside the disc $r<r_d=100$ and falls linearly to $0$ at $r=r_d+5$. The rate is $s=\min(0.5,\,2\pi A/M)$, where $M$ is the water in the disc. A full disc therefore loses the nominal volume $2\pi A$ per step and a nearly empty one loses less, so that the flux is selected by the flow.

On the sill, the drain is replaced by the bed profile $b(r)=b_0\exp[-(r-R_X)^2/W^2]$, with $b_0=0.7$ and $W=6$, and by a free overfall at the brink $r_{\rm brink}=R_X-15$. The bed enters the dynamics only through the conservative force $-gh\nabla b$. The cells at $r<r_{\rm brink}$ are not evolved: at each step their depth and swirl are copied from the ring just outside, and their inward velocity is set to the larger of the copied value and $1.05\sqrt{gh}$, so that nothing travels upstream from them. The brink is $15$ cells inside the crest, so the supercritical layer is $14$ cells wide. The sill flows are prepared in two stages. The drain flow is first settled for $32\,000$ steps; the drain is then switched off, the sill and the brink are switched on, and the flow is held for a further $12\,000$ steps, after which the depth changes by less than $10^{-4}$ over $1000$ steps \revr{(maximum absolute change of depth)}. \revr{Every flow is then held for a further $10^5$ steps before measurement, which removes a slow residual ringing.}

\revr{Waves are driven by rescaling the populations on a thin ring at every step, $f_i\to f_i[1+\epsilon\rho(r)\cos(m\theta-\omega t)]$, which modulates the depth by the same factor,} where $\rho$ is a Gaussian of width $2$ cells centred at $r=170$ and $\epsilon=2\times10^{-4}$. The fluxes are read on four rings, at $\rh+2$, $\rh+4$, $\rh+6$ and $\rh+8$, and averaged over four windows of the record. \revr{The modulation is switched on over $4000$ steps and held for a further $8000$; the following $16\,000$ steps are recorded in four windows.} Moving the drive to $r=140$ or $200$, or widening it to $4$ cells, changes the size of the fluxes but not their sign (Appendix~\ref{sm:ledger}).

\section{The lattice model}\label{sm:scheme}
The model follows Refs.~\cite{ShanYuanChen2006,Shan2010,Coreixas2017,Guo2002}: Hermite-expanded equilibria on Cartesian velocity sets, collisions in Hermite-moment space and second-order forcing.

\emph{Lattice and equilibrium.} The lattice is the tensor product of two one-dimensional five-speed sets $\{0,\pm1,\pm2\}$, with $c_s^2=1/2$ and weights $w=(9/16,5/24,5/24,1/96,1/96)$ per direction, so that $\mathbf e_i=(e_a,e_b)$ and $w_i=w_aw_b$. The equilibrium is
\begin{equation}
f_i^{\rm eq}=w_i\,h\prod_{\alpha=x,y}\;\sum_{n=0}^{4}\frac{H_n(e_{i\alpha})\,u_\alpha^n}{n!\,c_s^{2n}},
\label{eq:feq_sm}
\end{equation}
with $H_0=1$, $H_1=e$, $H_2=e^2-c_s^2$, $H_3=e^3-3c_s^2e$ and $H_4=e^4-6c_s^2e^2+3c_s^4$.

\emph{Collision and forcing.} The collision acts on the moments $m_{ab}=\sum_iH_a(e_{ix})H_b(e_{iy})f_i$ ($a,b=0,\dots,4$), each relaxing at its own rate,
\begin{equation}
m_{ab}^{\ast}=m_{ab}^{\rm eq}+(1-s_{ab})\,(m_{ab}-m_{ab}^{\rm eq})+(1-\tfrac12 s_{ab})\,S_{ab},
\label{eq:mrt_sm}
\end{equation}
with $s_{ab}=0$ for the conserved moments ($a+b\le1$), $s_{ab}=1/\tau$ for $a+b=2$ and $s_{ab}=s_h$ for $a+b\ge3$. $S_{ab}$ is the moment of the forcing term $S_i=w_i\big[(\mathbf e_i-\mathbf u)\cdot\mathbf F/c_s^2+(\mathbf e_i\cdot\mathbf u)(\mathbf e_i\cdot\mathbf F)/c_s^4\big]$ of Ref.~\cite{Guo2002}, with $\mathbf F=(c_s^2-gh)\nabla h-gh\nabla b$ and the half-step velocity correction. $\nabla h$ is the centred difference $[h(\mathbf x+\hat{\mathbf e})-h(\mathbf x-\hat{\mathbf e})]/2$ on the lattice; $\nabla b$ is the centred difference of the bed profile on the radial grid, interpolated to the cells. We use $\tau=0.70$ ($\nu=0.10$), $s_h=0.6$ and $g=0.05$.

\emph{Quadrature error.} The five-speed set reproduces the Gaussian moments up to the fourth, $\sum_iw_ie_i^4=3c_s^4$, but gives $\sum_iw_ie_i^6=7/4$ in place of $15c_s^6=15/8$. As a result $\sum_iw_ie_i^2H_4(e_i)=-1/8$, and the $n=4$ term of Eq.~\eqref{eq:feq_sm} adds $-hu_x^4/12$ per axis to the equilibrium momentum flux. This four-fold anisotropic error reaches a quarter of the hydrostatic pressure $gh^2/2$ in the supercritical layer ($h\simeq0.4$, $|u|\simeq0.43$) and stays below $0.3\%$ of it in the exterior ($h\simeq2$, $|u|\le0.2$). It belongs to the settled flow, is shared by the rotating hole and its twin, and is measured by the azimuthal spectrum of the settled depth (Appendix~\ref{sm:horizon}).

\emph{Dispersion.} The centred difference makes the wave speed depend on the wavelength $\lambda$. From the scheme linearised about a still layer, for waves along a lattice axis (along the diagonal the excess is about two thirds of these values), the excess of $c(k)$ over $\sqrt{gh}$ is:
\begin{center}
\begin{tabular}{lccc}
\hline
depth & $\lambda=32$ & $\lambda=21$ & $\lambda=16$ \\
\hline
$h=2$ (exterior) & $2\%$ & $4\%$ & $7\%$ \\
$h=1$ & $4\%$ & $10\%$ & $16\%$ \\
\revr{$h=0.7$ (crest, $R_X=115$)} & $7\%$ & $14\%$ & $23\%$ \\
\hline
\end{tabular}
\end{center}
 Waves longer than about $30$ cells travel at $\sqrt{gh}$ \revr{to better than $10\%$ at the depths of the exterior.} The driven modes, with azimuthal wavelengths of \revr{$100$--$180$} cells at the horizon, are in this regime. The sill ($W=6$) and the supercritical layer ($14$ cells) are shorter, so the one-way property of the layer applies to the long waves measured here.

\emph{Stability and checks.} With $\mathrm{Ma}=|\mathbf u|/c_s$, the flows of this paper reach $\mathrm{Ma}=0.6$ inside the horizon on the sills, held for $10^5$ steps, and $0.8$--$1.0$ on the drain.
\begin{center}
\begin{tabular}{p{0.54\columnwidth}p{0.40\columnwidth}}
\hline
check & result \\
\hline
von Neumann analysis, one-dimensional scheme, same equilibrium and rates & stable to $\mathrm{Ma}=1.30$; first instability at $1.35$ \\
pulses at $\mathrm{Ma}=0.95$, $1.15$, $1.30$ & bounded over $6000$ steps \\
uniform flow, every flow angle & fixed point to machine precision \\
shear viscosity against $c_s^2(\tau-\tfrac12)$ & agrees to $0.1\%$ \\
double-precision repeat (Appendix~\ref{sm:ledger}) & fluxes change by less than $10^{-4}$ \\
\hline
\end{tabular}
\end{center}

\emph{Bounds.} Two bounds keep $0.02\le h\le10$ and $|\mathbf u|\le1.6$ when $h$ and $\mathbf u$ are reconstructed from the populations. On the settled sill flows $h\ge0.39$ and $|\mathbf u|\le0.44$ everywhere, including inside the brink, so the bounds are never reached. On the drain the depth sits at the lower bound at $r<95$, inside the drain and behind a supercritical layer of about nine cells ($\Fr_r\ge1.2$); on the read-out rings the depth is $1.3$--$1.6$. The populations are stored in single precision; the double-precision repeat is in Appendix~\ref{sm:ledger}.

\section{Flow construction}\label{sm:flow}
 The box, the reservoir, the drain rule, the sill, the brink and the preparation of the flows are described in Appendix~\ref{app:methods}, and the parameters of the reference flows are listed in Table~\ref{tab:params}.

\emph{Drain.} The water $M$ in the disc, which sets the removal rate $s=\min(0.5,\,2\pi A/M)$, is counted with the depth floor $0.02$.

\emph{Sill.} The flow over the sill is choked. The sill height above which the crest becomes critical (the hydraulic-control threshold) is $\Delta b_c=E_{\rm up}/g-\tfrac32 h_c(q)-v_\theta^2(R_X)/2g$, evaluated with the realised flux $q$ and the energy of the exterior flow defined in Appendix~\ref{app:methods} (the Bernoulli head), swirl included:
\begin{center}
\begin{tabular}{lcccc}
\hline
sill & $q$ & $\Delta b_c$ & $b_0$ & crest depth against $h_c$ \\
\hline
$R_X=115$ & \revr{$14.5$} & $0.62$ & $0.7$ & \revr{within $3\%$} \\
$R_X=100$ & $10.8$ & $0.65$ & $0.7$ & \revr{within $7\%$} \\
\hline
\end{tabular}
\end{center}
On the non-rotating \revr{$R_X=115$} sill the crest depth equals the critical depth to $2\%$. Since $q$ is selected by the sill, this is a consistency check of the choked state.

\section{Outer boundary for perturbations}\label{sm:sponge}
The reservoir cells are reset at every step, so a wave reaching the edge of the disc is reflected. A sponge in the annulus $250<r<300$ absorbs outgoing waves. Once the flow has settled to $\mathbf u_{\rm ss}$, the body force
\begin{equation}
\mathbf F_{\rm s}=-\gamma_s\,w(r)^2\,h\,(\mathbf u-\mathbf u_{\rm ss}),\qquad \gamma_s=0.02,
\label{eq:sponge}
\end{equation}
is added there, with $w(r)$ rising linearly from $0$ to $1$ across the annulus. The force is not curl-free: it sheds vorticity, which the inflow carries to the drive radius, and this is why the sponge-on flux profiles reverse between $r=150$ and $165$ (Appendix~\ref{sm:ledger}). The sponge is therefore used only as a check.
\begin{center}
\begin{tabular}{p{0.60\columnwidth}p{0.32\columnwidth}}
\hline
check & result \\
\hline
undriven null test, change of depth over $1500$ steps & below $10^{-6}$ \\
reflectivity, $m=4$ and $6$, non-rotating $R_X=115$ sill & $2$--$6\%$ \\
reflectivity, $m=4$, rotating $R_X=115$ sill & $0.6$--$2.7\%$ \\
\hline
\end{tabular}
\end{center}
The reflectivity is the energy returning inward after the outgoing lobe has passed, for pulses of $6$ and $14$ steps launched at $r=200$ and read on rings at $r=210$--$240$; it is that of sponge and reservoir together. Waves long compared with the width of the sponge, such as the lowest tones of a ringdown, are partly returned~\cite{Flachi2026b}.

The sponge is used as follows:
\begin{center}
\begin{tabular}{p{0.34\columnwidth}cp{0.42\columnwidth}}
\hline
measurement & sponge & repeat \\
\hline
drain & on & $m=4$ without it (Table~\ref{tab:benchsink}) \\
sills, $m=4$ & off & every run with it (Table~\ref{tab:m4}) \\
sills, $m=6$ & off & co-rotating run on $R_X=115$ with it (Table~\ref{tab:m6}) \\
pulse and light-ring runs & on & -- \\
\hline
\end{tabular}
\end{center}

\section{Horizon formation}\label{sm:horizon}
\begin{figure*}[t]
\centering
\includegraphics[width=\textwidth]{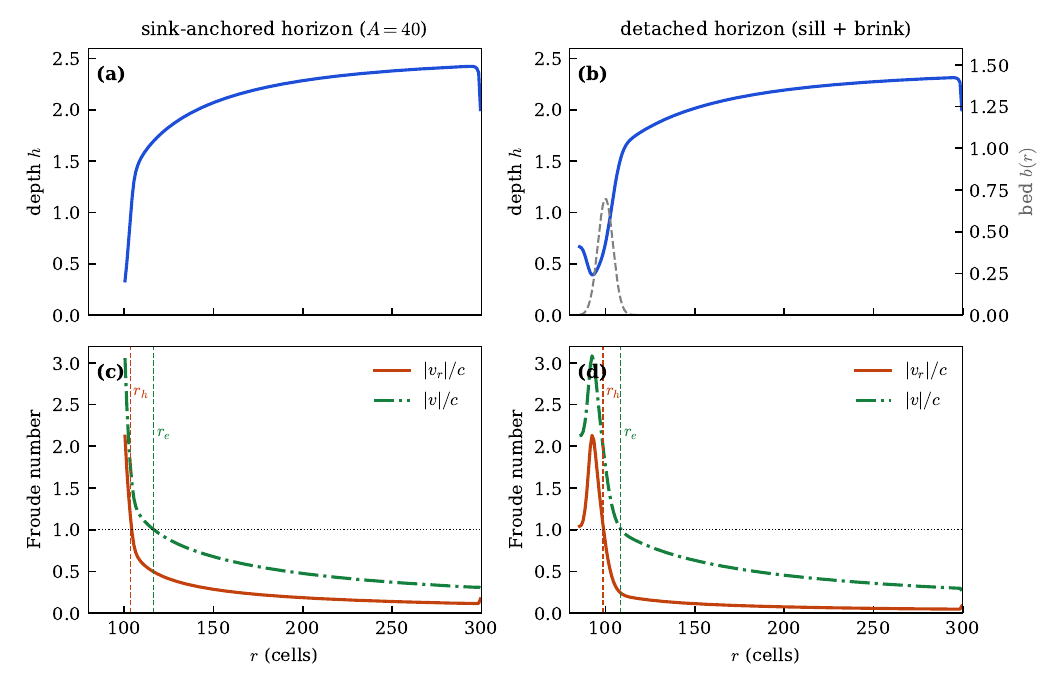}
\caption{\label{fig:bg}Settled flows: drain ($A=40$, left) and sill ($R_X=100$, right). (a),(b)~Depth, with the bed $b(r)$ dashed. (c),(d)~Radial and total Froude numbers; their unit crossings mark the horizon $\rh$ and the ergosurface $r_e$.}
\end{figure*}

Figure~\ref{fig:bg} shows the settled flows.

\emph{Horizon position.} The horizon is the unit crossing of $\Fr_r$, interpolated through the profiles. The ring value quoted elsewhere is the last one-cell shell with mean $\Fr_r\ge1$, labelled by its inner radius.
\begin{center}
\begin{tabular}{lccc}
\hline
flow & crossing & ring value & twin, ring value \\
\hline
sill $R_X=100$ & $99.0$ & $98$ & $99$ \\
sill $R_X=115$ & $114.3$ & $113$ & $114$ \\
drain $A=40$ & $104.3$ & -- & $104$ \\
\hline
\end{tabular}
\end{center}
On the sills the crossing lies $1.0$ and $0.7$ cells inside the crest. The read-out rings lie outside the twin's horizon in every case. In the supercritical layer $\Fr_r$ has a minimum of $1.04$, below the $1.05$ imposed at the brink, so the brink condition is active; a repeat with the brink moved is reported in Appendix~\ref{sm:ledger}.

\emph{Anisotropy.} The settled flows are axisymmetric up to the four-fold anisotropy of the lattice. The azimuthal Fourier amplitudes of the depth on the read-out rings are:
\begin{center}
\begin{tabular}{lc}
\hline
flow & $|h_4|/h_0$ \\
\hline
sill $R_X=115$ & $1.2\times10^{-3}$ \\
its twin & $0.9\times10^{-3}$ \\
sill $R_X=100$ & $1.0\times10^{-3}$ \\
drain & $5\times10^{-3}$ \\
\hline
\end{tabular}
\end{center}
In addition $|h_8|/h_0\le3\times10^{-4}$. This static pattern belongs to the background that is subtracted; it couples the driven $m=4$ wave to $m=0$ and $8$ at the $10^{-3}$ level.

\emph{Exterior of the drain and of the sill.} The two exteriors are compared on the settled flows ($A=40$ drain and $R_X=100$ sill, co-rotating) between $r=120$ and $250$, and through their light-ring frequencies, the frequencies of the peak response to a drive at $r=110$:
\begin{center}
\begin{tabular}{p{0.56\columnwidth}p{0.36\columnwidth}}
\hline
quantity & drain against sill \\
\hline
swirl $v_\theta$ & equal to $0.3\%$ \\
wave speed $\sqrt{gh}$ & equal to $1$--$2\%$ \\
depth & equal to $2$--$5\%$ \\
light-ring frequencies, $m=3$--$6$ (Table~\ref{tab:lr}) & equal to $2\%$ \\
radial inflow $|v_r|$ at $r=120$ & $0.14$ against $0.05$ \\
radial inflow $|v_r|$ at $r=250$ & $0.05$ against $0.02$ \\
\hline
\end{tabular}
\end{center}
The radial inflows are in the ratio of the two fluxes.

\section{Flux measurements}\label{sm:ledger}\label{sm:ledgers}
\begin{figure*}[t]\centering\includegraphics[width=0.98\textwidth]{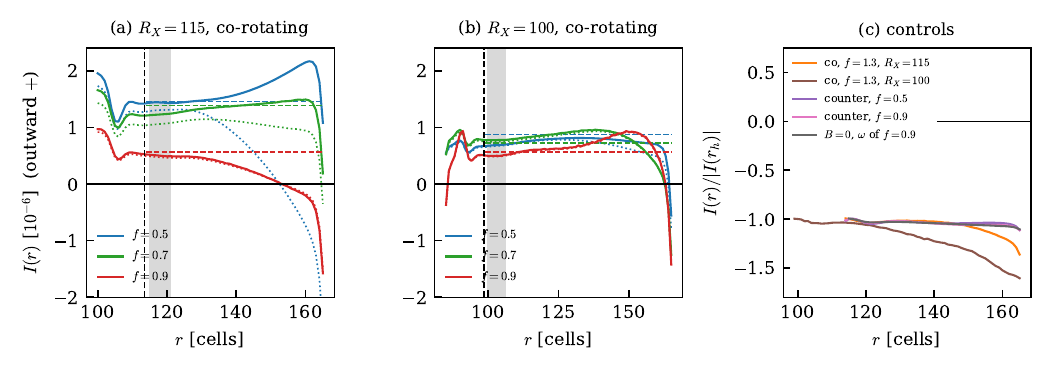}
\caption{\label{fig:prof}Energy flux $I(r)$ from the brink (left edge) through the horizon (dashed) to the drive, $m=4$. (a),(b)~Co-rotating wave in band: outward through the horizon. (c)~Controls (beyond the band edge, counter-rotating, non-rotating): inward. Grey: read-out rings. \revr{Dotted in (a),(b): the same runs with the outer sponge; in (b) they coincide with the solid curves up to $r=160$ at $f=0.7$ and $0.9$.} Dashed levels: the conserved flux of the regular continuum solution, Eq.~\eqref{eq:bench}, normalised to the lattice depth amplitude at $r=160$. All solid curves without the sponge.}
\end{figure*}

\emph{Protocol.} The drive is described in Appendix~\ref{app:methods}. The modulation is switched on over $4000$ steps and held for a further $8000$ steps; the following $16\,000$ steps are recorded in four consecutive windows. The observables are the time-averaged energy flux
\begin{equation}
I(r)=2\pi r\big\langle(g\,\delta h+\mathbf v_0\!\cdot\!\delta\mathbf u)(h_0\,\delta u_r+\delta h\,v_{0r})\big\rangle,
\label{eq:ledger}
\end{equation}
positive outward, and the angular-momentum flux $F_J(r)=2\pi r^2\langle(h_0\delta u_r+\delta h\,v_{0r})\delta u_\theta\rangle$, where the subscript $0$ denotes the settled flow and $\delta$ the departure from it. The net flux $I_{\rm net}$ is the mean of $I$ over the four rings $\rh+2$ to $\rh+8$. The counter-rotating run reverses the circulation, $B\to-B$. Every flow is held for $10^5$ steps beyond the settling criterion, which removes a slow residual ringing. Figure~\ref{fig:prof} shows the radial profile of the flux.

\emph{Floor and spreads.}
\begin{center}
\begin{tabular}{p{0.42\columnwidth}p{0.50\columnwidth}}
\hline
quantity & value \\
\hline
undriven floor & $\sim10^{-13}$, six to nine orders below the fluxes \\
acceptance & $|I_{\rm net}|$ above three times the floor \\
spread over the four windows & below $1\%$ \revr{($2.4\%$ for the $m=6$ flux at $f=1.0$ on $R_X=100$)} \\
spread over the four rings & $1.7$--$5.4\%$ (co-rotating), $3$--$6\%$ (twin) \\
outer-boundary protocol & \revr{in band up to $30\%$ (co-rotating) and $40\%$ (twin) over the three protocols; above the edge up to a factor $2$} \\
drive doubled & fluxes multiplied by $4.000$ (co-rotating at $f=0.5$, $0.9$, $1.3$; twin at $f=0.5$) \\
\hline
\end{tabular}
\end{center}

\emph{Lowest frequency.} At $f=0.3$ the turning radius lies far outside the box, and the twin's flux depends on the boundary protocol (Table~\ref{tab:m4}). The twin is therefore used, and $G$ quoted, from $f=0.5$ upwards, \revr{where the sponge changes its flux by less than $25\%$ in band \revr{(up to a factor $2$ above the edge)} and changes no sign.}

\emph{Band edge.} The band edge moves with $\OmH$ and with $m$, which excludes a resonance of the box. On $R_X=115$ the $m=4$ wave is absorbed at $f=1.1$ and $1.15$ ($I_{\rm net}=-1.1$ and $-2.3\times10^{-6}$), at absolute frequencies that are still in band on $R_X=100$. On $R_X=100$ the $m=4$ flux changes sign between $f=0.95$ and $1.0$ (sponge on, units of $10^{-6}$):
\begin{center}
\begin{tabular}{lccccc}
\hline
$f$ & $0.95$ & $1.0$ & $1.05$ & $1.1$ & $1.2$ \\
\hline
$I_{\rm net}$ & $+0.32$ & $-0.13$ & $-1.49$ & $-5.38$ & $-7.44$ \\
\hline
\end{tabular}
\end{center}

\emph{Checks.} On the $R_X=115$ sill, $m=4$, sponge on unless stated:
\begin{center}
\begin{tabular}{p{0.40\columnwidth}p{0.52\columnwidth}}
\hline
check & result \\
\hline
double precision & fluxes change by less than $10^{-4}$; for $m=6$ at $f=0.9$--$1.1$ on both sills (no sponge), equal to three figures \\
brink moved from $r=100$ to $97$ or $103$ & $I^{\rm co}_{\rm net}$ at $f=0.5$ changes by less than $0.5\%$, $I^{B=0}_{\rm net}$ by less than $0.3\%$; inward flux at $f=1.3$ by $2\%$ \\
drive moved to $r=140$ or $200$, or widened to $4$ cells & flux multiplied by $2.9$, $0.55$ and $4.0$ at $f=0.5$; signs unchanged at $f=0.5$ and $0.9$ \\
every length doubled & $G=0.31$ at $f=0.5$, as at the original size with the sponge; flux flat to $1\%$ over the rings \\
perturbation launched inside the supercritical layer & flux outside the horizon $10^{-4}$ to $10^{-6}$, and depth amplitude $0.2$--$1.5\%$, of those for the same perturbation launched outside \\
\hline
\end{tabular}
\end{center}
The doubled run has $R_X=230$, brink at $200$, $W=12$, drive at $340$, free disc to $600$, sponge in $500$--$600$, $N=1536$, $A=40$ and $B=60$, giving $\rh=228$ and $\OmH=1.133\times10^{-3}$. The one-way test uses depth bumps of width $1.5$ or $3$ cells with $m=0$ or $4$, launched at $r=106$ (inside the layer) and at $r=128$ (outside) and read on rings at $r=118$--$145$, on the rotating and the non-rotating flow; the undriven floor is three orders of magnitude lower.

\emph{Other azimuthal numbers.} On $R_X=115$, for $m=2$ (sponge on, units of $10^{-6}$):
\begin{center}
\begin{tabular}{lccccc}
\hline
$f$ & $0.3$ & $0.5$ & $0.7$ & $0.9$ & $1.3$ \\
\hline
co-rotating & $+12.2$ & $+14.3$ & $+6.0$ & $+10.7$ & $-3.6$ \\
twin & $-1.2$ & $-43.1$ & $-15.1$ & $-9.5$ & $-50.9$ \\
\hline
\end{tabular}
\end{center}
so that $G=0.33$ at $f=0.5$. For $m=3$ and $5$ the co-rotating flux is outward at $f=0.3$--$0.9$ and inward at $f=1.3$, the twins absorb at every frequency, and $G=0.15$ and $0.45$ at $f=0.5$.

\emph{Data tables.} Tables~\ref{tab:m4} and \ref{tab:m6} give the fluxes on the sills for $m=4$ and $6$, and Table~\ref{tab:m6x100} the $m=6$ fluxes on $R_X=100$. Table~\ref{tab:sink} gives those on the drain: $A=40$ with its twin; $A=20$, on which the reservoir is matched ($q=19.2$); and $A=16$ and $14$, with $r_d=80$ and $70$, $\rh=83$ and $73$, and $\OmH=4.2$ and $5.2\times10^{-3}$. Table~\ref{tab:visc} gives the $R_X=115$ sill at lower viscosity.

\emph{Flux away from the horizon.} Between the read-out rings and the drive the flux varies with radius (Fig.~\ref{fig:prof}) and depends on the treatment of the outer boundary. On $R_X=115$ at $f=0.5$ and $\nu=0.10$ (units of $10^{-6}$):
\begin{center}
\begin{tabular}{lcccc}
\hline
protocol & $I^{\rm co}_{\rm net}$ & $I^{B=0}_{\rm net}$ & $G$ & $I^{\rm co}(160)/I^{\rm co}_{\rm net}$ \\
\hline
no sponge & $+1.44$ & $-3.49$ & $0.41$ & $1.5$ \\
sponge on the velocity & $+1.30$ & $-4.19$ & $0.31$ & $-0.56$ \\
sponge on the depth only & $+1.11$ & $-2.98$ & $0.37$ & $1.8$ \\
\hline
\end{tabular}
\end{center}
Without the sponge the flux stays outward to $r=150$ at every in-band frequency on both sills; the conserved flux of the regular continuum solution, normalised at $r=160$, lies at the level of these profiles (dashed in Fig.~\ref{fig:prof}). At $f=0.9$ the profile reverses at $r\approx153$ in all three protocols. The spread of $G$ over the three protocols, $0.31$--$0.41$, is the systematic quoted on it.

$G$ keeps its sign at the three viscosities under both protocols (Table~\ref{tab:visc}). At $\nu=0.025$ and $f=0.5$ the absolute fluxes depend on the protocol.

\emph{Energy budget.} {With $p=g\,\delta h+\mathbf u_0\cdot\delta\mathbf u$ and $\mathbf q=h_0\,\delta\mathbf u+\delta h\,\mathbf u_0$, the linearised lattice equations about the steady background give exactly
\begin{equation}
\partial_t E+\nabla\cdot(p\,\mathbf q)=p\,s+\mathbf q\cdot\delta\mathbf a-h_0\,\delta\omega\,\delta\mathbf u\cdot\mathbf u_0^{\perp}-\omega_0\,\delta h\,\mathbf u_0\cdot\delta\mathbf u^{\perp},
\label{eq:budget}
\end{equation}
where $E=\tfrac12h_0|\delta\mathbf u|^2+\delta h\,\mathbf u_0\cdot\delta\mathbf u+\tfrac12g\,\delta h^2$, $s$ is the mass injected by the drive, $\delta\mathbf a$ is the non-inviscid acceleration of the perturbation (viscous stress, sponge and lattice corrections, measured from the fields), $\omega=\nabla\times\mathbf u$ and $\mathbf u^{\perp}=(-u_y,u_x)$. Evaluated between the read-out rings and $r=160$ on $R_X=115$ ($m=4$, sponge on), Eq.~(\ref{eq:budget}) closes to $0.1\%$, $1\%$ and $9\%$ at $f=0.5$, $0.9$ and $1.3$ , and to $2\%$ on the $B=0$ twin. The change of the flux is carried mostly by the last two terms, the exchange of energy between the wave and the vorticity of the perturbation and of the background: $97\%$ of it at $f=0.5$, $76\%$ at $f=0.9$ and all of it on the twin; at $f=1.3$ it is mostly viscous work. A budget with the viscous and bulk dissipation alone misses this exchange: it accounts for one third of the change at $f=0.5$ and has the wrong sign on the twin.} The change does not scale with the viscosity (Table~\ref{tab:visc}). At $f=0.5$ the perturbation at $r=160$ is $80$--$90\%$ vortical by the measure $|\nabla\times\delta\mathbf u|^2/(|\nabla\times\delta\mathbf u|^2+|\nabla\cdot\delta\mathbf u|^2)$, against $2\%$ next to the horizon.

\emph{Checks at the rings.} For a single $(m,\omega)$ mode $F_J=(m/\omega)\,I$. Outside $0.9\le f\le1.0$, where $I$ passes through zero, the measured $F_J/I$ agrees with $m/\omega$ (Table~\ref{tab:fj}). The flux continues through the horizon into the supercritical layer (Fig.~\ref{fig:prof}).
The exterior swirl is a potential vortex to $2$--$3\%$.

\section{Near-field gain from the net flux}\label{sm:gain}
\emph{Turning radius.} In band the wave is evanescent between its inner turning point, a few cells outside the horizon, and the turning radius
\begin{equation*}
r_t=\frac{1}{2f\OmH}\Big[c+\sqrt{c^2+4f\OmH\Gamma}\Big],
\end{equation*}
with $\Gamma=v_\theta r$ and $c$ of the exterior, and travels only beyond it. For $f=0.3$--$0.9$:
\begin{center}
\begin{tabular}{lc}
\hline
flow & $r_t$ (cells) \\
\hline
sills & $185$--$560$ \\
drain $A=40$ & $190$--$460$ \\
drain $A=16$ & $140$--$324$ \\
drain $A=14$ & $121$--$273$ \\
\hline
\end{tabular}
\end{center}
With the drive at $r=170$, there is no incident travelling wave to normalise by on the sills.

\emph{Definition.} For equal drive, the gain
\begin{equation}
G=\frac{I^{\rm co}_{\rm net}}{|I^{B=0}_{\rm net}|}
\label{eq:gain}
\end{equation}
is the energy returned by the rotating horizon in units of the energy its twin absorbs. Its size depends on the twin, which is driven closer to its travelling regime than the rotating hole (WKB transparency ratios $T_{\rm co}/T_{B=0}$, Table~\ref{tab:gain}) and whose flux depends on the outer boundary. Table~\ref{tab:gain} lists both; their difference is the boundary systematic on $G$, quoted in the text. The amplitude-normalised gain $\hat G$ of Appendix~\ref{sm:bench} does not depend on the boundary.

\emph{Absorption by the twin.}\label{sm:sillpulse}\label{sm:refl} A pulse of width $6$, launched at $r=150$ onto the non-rotating $R_X=115$ sill and read on rings at $r=118$--$145$, returns at most the following fraction of its energy:
\begin{center}
\begin{tabular}{lccc}
\hline
$m$ & $0$ & $2$ & $4$ \\
\hline
returned energy & $1.4\%$ & $0.4\%$ & $1.0\%$ \\
\hline
\end{tabular}
\end{center}
The twin's horizon absorbs what reaches it.

\emph{Rotating flow and twin.} The two flows differ next to the horizon (rotating / twin):
\begin{center}
\begin{tabular}{lcc}
\hline
quantity & $R_X=115$ & $R_X=100$ \\
\hline
flux $q$ & $14.5$ / $21.7$ & $10.8$ / $20.1$ \\
\hline
\end{tabular}
\end{center}
The twin's layer is deeper by $20$--$80\%$ within eight cells of the horizon. The bars of Fig.~\ref{fig:gain} combine the window spread with a $15\%$ systematic for this mismatch, or half the sponge-on/off difference where that is larger, which happens only above the band edge. The two exteriors agree to within $10\%$ beyond thirty cells.

\emph{Drain.} On the $A=40$ drain $G=0.15$ ($m=4$) and $0.17$ ($m=6$) at $f=0.5$ (Table~\ref{tab:sink}). \revr{At $f=0.5$ and $1.3$ the signs are unchanged under the sponge protocol, the viscosity and the resolution; at $f=0.9$ the co-rotating flux reverses at lower viscosity (Appendix~\ref{sm:bench}).} 

\emph{Larger box.}\label{sm:bigbox} A direct reading of $|R|^2$ requires a box in which the in-band wave travels. The $R_X=115$ sill was rebuilt in a box five times larger ($N=3072$, $r_{\rm out}=1500$, drive at $r=900$). Near the horizon the fluxes have the signs of the small box (units of $10^{-6}$):
\begin{center}
\begin{tabular}{lcccc}
\hline
$f$ & $0.5$ & $0.7$ & $0.9$ & $1.3$ \\
\hline
co-rotating & $+0.14$ & $+0.73$ & $+0.84$ & $-17.6$ \\
twin & $-1.7$ & $-29$ & $-104$ & $-126$ \\
\hline
\end{tabular}
\end{center}
In the travelling region the horizon's contribution is about $1\%$, against a viscous attenuation of about $30\%$, and $|R|^2$ cannot be extracted: the in/out decomposition, calibrated on the outward emission of the drive, returns values between $0.08$ and $0.95$ for the co-rotating wave and between $-0.08$ and $0.97$ for the twin.

\section{Continuum benchmark on the measured background}\label{sm:bench}
\emph{Equation.} The linearised shallow-water equations for potential perturbations, $\delta\mathbf u=\nabla\phi$ and $\delta h=-(\partial_t+\mathbf v_0\!\cdot\!\nabla)\phi/g$, on the measured axisymmetric flow $h_0(r)$, $v_r(r)$, $v_\theta(r)$ reduce, for $\phi(r)e^{i(m\theta-\omega t)}$, to
\begin{align}
\mathcal A\phi''+\mathcal B\phi'+\mathcal C\phi&=0,\qquad \mathcal A=v_r^2-c^2,\nonumber\\
\mathcal B&=-2i\varpi v_r+\frac{[r(v_r^2-c^2)]'}{r},\nonumber\\
\mathcal C&=-\varpi^2-\frac{i}{r}(rv_r\varpi)'+\frac{m^2c^2}{r^2},
\label{eq:bench}
\end{align}
with $c^2=gh_0$ and $\varpi=\omega-mv_\theta/r$. $\mathcal A$ vanishes at the horizon. The solution regular there ($\mathcal B\phi'+\mathcal C\phi=0$ at $\rh$) is integrated outward to the drive; its flux $I=\pi r\,{\rm Re}[\,i\omega\phi\,(h_0\phi'+\delta h\,v_r)^\ast]$ is conserved to $10^{-6}$. The regular solution carries an outward flux in band and an inward flux above the edge on the rotating flows, and an inward flux at every frequency on the twins.

\emph{Comparison with the lattice.} The comparison has no free parameter. The lattice depth amplitude $|C(r)|$ is the modulus of the complex amplitude of $\delta h$ at the drive frequency and azimuthal number, accumulated over the measurement windows. The amplitude-normalised gain is $\hat G=[I^{\rm co}_{\rm net}/|C^{\rm co}(160)|^2]\,/\,[|I^{B=0}_{\rm net}|/|C^{B=0}(160)|^2]$, predicted by solving Eq.~\eqref{eq:bench} on both flows.
\begin{center}
\begin{tabular}{p{0.50\columnwidth}p{0.42\columnwidth}}
\hline
quantity & lattice against continuum \\
\hline
radial profile of the depth amplitude (Fig.~\ref{fig:bench}) & \revr{within $5\%$ at $f=0.5$ and $1.3$ and $14\%$ at $f=0.7$ between the read-out rings and $r=160$; at $f=0.9$ the lattice amplitude is up to $52\%$ higher at $r=160$} \\
flux next to the horizon per unit depth amplitude at $r=160$ (Table~\ref{tab:bench}) & ratio $0.8$--$1.1$ at every $f$ on both sills \\
$\hat G$ (Table~\ref{tab:bench}) & agree to $2$--$20\%$, sign included \\
\hline
\end{tabular}
\end{center}
 The equal-drive gain $G$ of Appendix~\ref{sm:gain} is not predicted by the regular solution, because the depth amplitude that each flow develops at $r=160$ depends on the response of the whole box (Table~\ref{tab:gain}).

 $\OmH$ is read at the centre of the last supercritical shell ($r=113.5$ and $98.5$); at the crossings it is lower by $1.3\%$ and $1.0\%$, as for a potential vortex, a shift common to every $f$ quoted.

\begin{figure*}[t]\centering\includegraphics[width=0.98\textwidth]{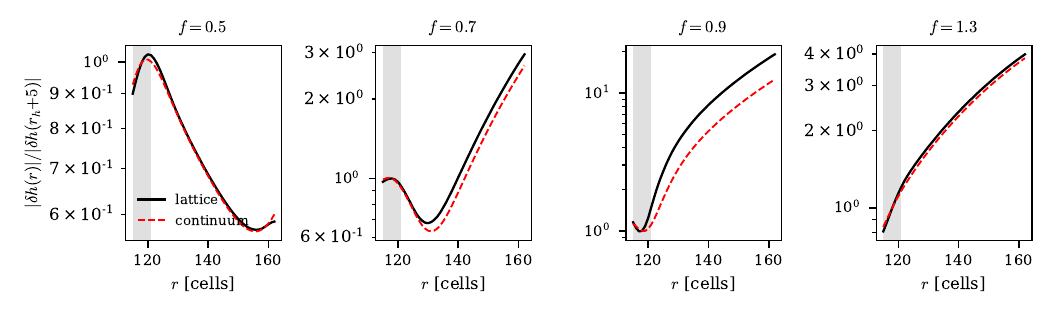}
\caption{\label{fig:bench}Depth amplitude of the driven $m=4$ wave on the $R_X=115$ sill (no sponge), normalised at $\rh+5$: lattice lock-in $|C(r)|$ (black) and the continuum solution of Eq.~\eqref{eq:bench} regular at the horizon (red dashed). Grey: read-out rings. The ratio of their end values at the drive side sets the comparison of Table~\ref{tab:bench}.}
\end{figure*}

\emph{Three viscosities.} The benchmark is repeated at $\nu=0.10$, $0.05$ and $0.025$ on $R_X=115$, without the sponge (Table~\ref{tab:benchnu}). The regular solution is inviscid and changes with $\nu$ only through the flow, whereas a viscous absorber would return a flux that scales with $\nu$. The flux is normalised at $\rh+5$, where the perturbation is $98\%$ potential.

At $f=0.5$ the lattice-to-continuum ratio decreases together for the co-rotating wave and for the twin, and their ratio $\hat G$ agrees with the prediction to within $7\%$ at each viscosity. At $f=0.9$ the co-rotating ratio falls to $1$ as $\nu$ decreases, and the twin's stays at $1$.

Normalised at $r=160$, the $f=0.5$ ratios fall with $\nu$; the field there is $80$--$90\%$ vortical (Appendix~\ref{sm:ledger}), which the potential equation does not describe. Also at $f=0.9$:
\begin{center}
\begin{tabular}{p{0.50\columnwidth}ccc}
\hline
$\nu$ & $0.10$ & $0.05$ & $0.025$ \\
\hline
lattice amplitude above continuum at $r=160$ & $52\%$ & $36\%$ & $13\%$ \\
flux per unit depth amplitude at $r=160$, co-rotating & $0.39$ & $0.36$ & $0.40$ \\
flux per unit depth amplitude at $r=160$, twin & $2.11$ & $2.15$ & $2.16$ \\
\hline
\end{tabular}
\end{center}

\emph{Drain.} The same checks on the drain ($A=40$, $r_d=100$, $m=4$) are in Table~\ref{tab:benchsink}. At $f=0.5$ the co-rotating flux is outward and the twin's inward at the three viscosities, with and without the sponge, and on a drain of twice the size ($A=80$, $r_d=200$, $N=1536$, $\rh=203$, $\OmH=1.44\times10^{-3}$). At $f=1.3$ the flux is inward in every run. At $f=0.9$ the co-rotating flux is outward at $\nu=0.10$ and reverses at $\nu=0.05$ and $0.025$.

Per unit depth amplitude at $\rh+5$, the twin absorbs two to three times the inviscid horizon flux at the three viscosities, a factor that does not decrease with $\nu$. The removal region spans $r=100$ to $105$ and acts on the subcritical cells just outside the horizon ($\Fr_r=0.72$--$0.82$), where it removes wave energy in proportion to the local amplitude squared, the same scaling as the horizon's own extraction. On the sill the brink lies fourteen cells behind the horizon, at $\Fr_r>1$ throughout, and nothing returns from it (Appendix~\ref{sm:ledger}).

\FloatBarrier

\clearpage

\begin{table*}[!ht]
\caption{\label{tab:params}Parameters of the reference flows. Lengths in cells, times in steps, speeds in lattice units. Common to both: D2Q25 lattice with $c_s^2=1/2$; $\tau=0.70$ ($\nu=0.10$); $g=0.05$, $h_{\rm res}=2$, $c_{\rm res}=0.316$; $N=768$, $r_{\rm out}=300$; $B=30$; sponge with $\gamma_s=0.02$ in $250<r<300$.}
\begin{ruledtabular}
\begin{tabular}{p{0.30\textwidth}p{0.32\textwidth}p{0.32\textwidth}}
 & drain & sill ($R_X=100$)\\
\colrule
$A$ & $40$ & $20$\\
exit & removal in the disc $r_d=100$ & sill $b_0=0.7$, $W=6$; brink at $r=85$\\
realised flux $q$ & $28.5$ & $10.8$\\
horizon, crossing & $104.3$ & $99.0$\\
$\OmH$ & $2.77\times10^{-3}$ & $2.98\times10^{-3}$\\
ergoregion & $\rh<r\lesssim116$ & $\rh<r\le108$\\
depth ratio $h(r{=}250)/h(\rh)$ & $2.1$ & $3.9$\\
twin ($B=0$): horizon, ring value & $104$ & $99$\\
twin ($B=0$): realised flux $q$ & --- & $20.1$\\
between horizon and exit & removal region at $r=100$--$105$, across the horizon & supercritical layer, $14$ cells ($\Fr_r\ge1.04$)\\
\revr{maximum change of depth over $1000$ steps after settling} & $10^{-6}$ & $10^{-5}$\\
\end{tabular}
\end{ruledtabular}
\end{table*}

\begin{table}[!ht]
\caption{\label{tab:sink}Net energy flux on the drain, in units of $10^{-6}$: mean over the rings $r_d+6$ to $r_d+12$ and the four windows ($B=30$, sponge on). $A=40$ is the drain of the main text, $A=20$ has the reservoir matched, and $A=16$ and $14$ form the rotation scan. co: co-rotating; ctr: counter-rotating; $B=0$: the twin of $A=40$ at the same $\omega$. The counter-rotating and $B=0$ fluxes at $f=1.1$ and $1.3$ are close but come from separate runs.}
\begin{ruledtabular}
\begin{tabular}{lcccccc}
$m=4$, $f$ & $0.3$ & $0.5$ & $0.7$ & $0.9$ & $1.1$ & $1.3$ \\ \colrule
$A=40$ co & $+1.43$ & $+0.88$ & $+0.21$ & $+0.27$ & $-4.73$ & $-18.7$ \\
$A=40$ ctr & $-27.0$ & $-44.2$ & $-72.2$ & $-31.1$ & $-22.5$ & $-54.1$ \\
$A=40$, $B=0$ & $-1.72$ & $-5.80$ & $-31.9$ & $-73.5$ & $-22.6$ & $-55.2$ \\
$A=20$ co & $+1.35$ & $+1.37$ & $+0.84$ & $+0.59$ & --- & $-5.23$ \\
$A=20$ ctr & $-24.8$ & $-42.3$ & $-37.7$ & $-11.9$ & --- & $-35.2$ \\
$A=16$ co & $+0.53$ & $+0.50$ & $+1.31$ & $+0.37$ & $-7.79$ & $-24.6$ \\
$A=16$ ctr & $-28.3$ & $-41.2$ & $-16.0$ & $-38.6$ & $-16.7$ & $-29.1$ \\
$A=14$ co & $+0.37$ & $+0.41$ & $+0.99$ & $+2.06$ & $-4.58$ & $-28.3$ \\ \colrule
$m=6$, $f$ & $0.5$ & $0.8$ & $0.95$ & $1.1$ & $1.25$ & \\ \colrule
$A=40$ co & $+0.136$ & $+0.120$ & $+0.132$ & $-0.435$ & $-5.26$ & \\
$A=40$ ctr & $-62.6$ & $-35.4$ & $-47.8$ & $-24.9$ & $-33.1$ & \\
$A=40$, $B=0$ & $-0.80$ & $-23.1$ & $-60.3$ & $-48.3$ & $-34.8$ & \\
$A=20$ co & $+0.101$ & $+0.168$ & $+0.223$ & $-0.655$ & $-11.8$ & \\
$A=20$ ctr & $-27.4$ & $-33.6$ & $-18.5$ & $-22.0$ & $-25.8$ & \\
$A=16$ co & $+0.037$ & $+0.154$ & $+0.085$ & $-1.85$ & $-9.06$ & \\
$A=16$ ctr & $-20.5$ & $-19.5$ & $-17.3$ & $-16.7$ & $-14.1$ & \\
$A=14$ co & $+0.046$ & $+0.144$ & $+0.155$ & $-2.00$ & $-12.6$ & \\
\end{tabular}
\end{ruledtabular}
\end{table}

\begin{table*}[!ht]
\caption{\label{tab:visc}The $R_X=115$ sill at three viscosities ($m=4$, fluxes in units of $10^{-6}$, read-out as in Table~\ref{tab:m4}), with the sponge on and off. Last column: co-rotating flux at $r=160$ relative to $I^{\rm co}_{\rm net}$, sponge on.}
\begin{ruledtabular}
\begin{tabular}{ccccccccc}
$\nu$ ($\tau$) & $\rh$ & $\OmH\,[10^{-3}]$ & $f$ & $I^{\rm co}_{\rm net}$, $I^{B=0}_{\rm net}$ & $G$ & $I^{\rm co}_{\rm net}$, $I^{B=0}_{\rm net}$ (off) & $G$ (off) & $I(160)/I_{\rm net}$ \\ \colrule
$0.10$ ($0.70$) & $113$ & $2.270$ & $0.5$ & $+1.30$, $-4.19$ & $0.31$ & $+1.44$, $-3.49$ & $0.41$ & $-0.56$ \\
 & & & $0.9$ & $+0.47$, $-53.9$ & $0.009$ & $+0.50$, $-58.3$ & $0.009$ & $-0.37$ \\
$0.05$ ($0.60$) & $113$ & $2.293$ & $0.5$ & $+2.13$, $-5.30$ & $0.40$ & $+1.18$, $-2.70$ & $0.44$ & $-0.80$ \\
 & & & $0.9$ & $+0.46$, $-54.9$ & $0.008$ & $+0.48$, $-65.1$ & $0.007$ & $-1.10$ \\
$0.025$ ($0.55$) & $114$ & $2.264$ & $0.5$ & $+2.76$, $-6.40$ & $0.43$ & $+0.82$, $-1.68$ & $0.49$ & $-1.12$ \\
 & & & $0.9$ & $+0.66$, $-56.6$ & $0.012$ & $+0.51$, $-60.3$ & \revr{$0.008$} & $-0.24$ \\
\end{tabular}
\end{ruledtabular}
\end{table*}

\begin{table*}[!ht]
\caption{\label{tab:m4}Net energy flux $I_{\rm net}$ for the $m=4$ wave against $f=\omega/m\OmH$, in units of $10^{-6}$, outward positive: mean over the four read-out rings and the four windows. Co-rotating and counter-rotating runs on the two sills, and the $B=0$ run of each sill at the same absolute frequencies. Upper block: with the sponge. Lower block: without it (used in the main text). The spread over the four windows is at most $0.03$ in these units.}
\begin{ruledtabular}
\begin{tabular}{lccccc}
run & $f=0.3$ & $f=0.5$ & $f=0.7$ & $f=0.9$ & $f=1.3$ \\ \colrule
\multicolumn{6}{l}{\emph{with sponge}}\\
co-rotating, $R_X=100$ & $+0.818$ & $+0.670$ & $+0.747$ & $+0.488$ & $-3.936$ \\
co-rotating, $R_X=115$ & $+1.787$ & $+1.302$ & $+1.060$ & $+0.473$ & $-14.159$ \\
counter-rotating, $R_X=100$ & $-29.59$ & $-46.19$ & $-62.38$ & $-16.780$ & $-46.066$ \\
counter-rotating, $R_X=115$ & $-22.90$ & $-28.829$ & $-53.343$ & $-52.900$ & $-18.732$ \\
$B=0$, $R_X=100$ & $-0.379$ & $-2.703$ & $-29.60$ & $-49.85$ & $-57.66$ \\
$B=0$, $R_X=115$ & $-1.117$ & $-4.186$ & $-9.486$ & $-53.92$ & $-24.168$ \\
\colrule
\multicolumn{6}{l}{\emph{without sponge}}\\
co-rotating, $R_X=100$ & $+0.762$ & $+0.686$ & $+0.775$ & $+0.500$ & $-1.948$ \\
co-rotating, $R_X=115$ & $+1.652$ & $+1.435$ & $+1.228$ & $+0.501$ & $-16.690$ \\
counter-rotating, $R_X=100$ & $-42.57$ & $-43.39$ & $-65.57$ & $-10.936$ & $-53.331$ \\
counter-rotating, $R_X=115$ & $-26.91$ & $-24.287$ & $-54.719$ & $-54.80$ & $-22.649$ \\
$B=0$, $R_X=100$ & $-0.536$ & $-2.810$ & $-31.83$ & $-40.20$ & $-79.94$ \\
$B=0$, $R_X=115$ & $-0.517$ & $-3.491$ & $-9.060$ & $-58.33$ & $-14.531$ \\
\end{tabular}
\end{ruledtabular}
\end{table*}

\begin{table*}[!ht]
\caption{\label{tab:m6}As Table~\ref{tab:m4}, for the $m=6$ wave on $R_X=115$ without the sponge: co-rotating run, its repeat with the sponge, counter-rotating run and $B=0$ run. The spread over the four windows is at most $0.009$.}
\begin{ruledtabular}
\begin{tabular}{ccccc}
$f$ & co-rot. & co-rot., sponge & counter-rot. & $B=0$ \\ \colrule
$0.20$ & $+0.340$ & $+0.348$ & $-4.345$ & $-0.048$ \\
$0.35$ & $+0.321$ & $+0.373$ & $-36.315$ & $-0.221$ \\
$0.50$ & $+0.292$ & $+0.314$ & $-43.798$ & $-0.898$ \\
$0.65$ & $+0.239$ & $+0.240$ & $-46.166$ & $-3.425$ \\
$0.80$ & $+0.216$ & $+0.213$ & $-11.453$ & $-29.619$ \\
$0.90$ & $+0.179$ & $+0.181$ & $-27.491$ & $-41.526$ \\
$1.00$ & $+0.037$ & $+0.037$ & $-52.215$ & $-11.928$ \\
$1.10$ & $-0.546$ & $-0.514$ & $-45.992$ & $-53.862$ \\
$1.25$ & $-3.065$ & $-3.603$ & $-15.323$ & $-75.344$ \\
$1.50$ & $-30.327$ & $-18.715$ & --- & $-20.283$ \\
\end{tabular}
\end{ruledtabular}
\end{table*}

\begin{table}[!ht]
\caption{\label{tab:m6x100}As Table~\ref{tab:m4}, for the $m=6$ wave on $R_X=100$ without the sponge. The spread over the four windows is at most $0.008$.}
\begin{ruledtabular}
\begin{tabular}{cccc}
$f$ & co-rot. & counter-rot. & $B=0$ \\ \colrule
$0.20$ & $+0.136$ & $-6.136$ & $-0.022$ \\
$0.35$ & $+0.113$ & $-35.025$ & $-0.105$ \\
$0.50$ & $+0.087$ & $-53.697$ & $-0.680$ \\
$0.65$ & $+0.117$ & $-12.667$ & $-17.759$ \\
$0.80$ & $+0.111$ & $-47.963$ & $-9.477$ \\
$0.90$ & $+0.177$ & $-39.713$ & $-75.530$ \\
$1.00$ & $-0.030$ & $-14.178$ & $-59.304$ \\
$1.10$ & $-0.276$ & $-18.395$ & $-26.235$ \\
$1.25$ & $-12.312$ & $-43.208$ & $-46.697$ \\
$1.50$ & $-27.377$ & $-14.331$ & $-24.100$ \\
\end{tabular}
\end{ruledtabular}
\end{table}

\begin{table*}[!ht]
\caption{\label{tab:gain}Equal-drive gain $G=I^{\rm co}_{\rm net}/|I^{B=0}_{\rm net}|$ with the sponge on and off, from the fluxes of Tables~\ref{tab:m4}--\ref{tab:m6x100}, and the WKB ratio $T_{\rm co}/T_{B=0}$ of the barrier transparencies of the rotating flow and its twin. The $m=6$ twins were run without the sponge only. $f=0.3$ is omitted: there the twin's flux is set by the box.}
\begin{ruledtabular}
\begin{tabular}{ccccc}
sill, $m$ & $f$ & $G$ (sponge) & $G$ (no sponge) & $T_{\rm co}/T_{B=0}$ \\ \colrule
$R_X=115$, $4$ & $0.50$ & $+0.311$ & $+0.411$ & $0.65$ \\
$R_X=115$, $4$ & $0.70$ & $+0.112$ & $+0.136$ & $0.35$ \\
$R_X=115$, $4$ & $0.90$ & $+0.009$ & $+0.009$ & $0.12$ \\
$R_X=115$, $4$ & $1.30$ & $-0.586$ & $-1.149$ & $0.13$ \\
$R_X=100$, $4$ & $0.50$ & $+0.248$ & $+0.244$ & $0.40$ \\
$R_X=100$, $4$ & $0.70$ & $+0.025$ & $+0.024$ & $0.09$ \\
$R_X=100$, $4$ & $0.90$ & $+0.010$ & $+0.012$ & $0.03$ \\
$R_X=100$, $4$ & $1.30$ & $-0.068$ & $-0.024$ & $0.20$ \\
$R_X=115$, $6$ & $0.50$ & --- & $+0.325$ & $0.52$ \\
$R_X=115$, $6$ & $0.65$ & --- & $+0.070$ & $0.27$ \\
$R_X=115$, $6$ & $0.80$ & --- & $+0.007$ & $0.10$ \\
$R_X=115$, $6$ & $0.90$ & --- & $+0.004$ & $0.04$ \\
$R_X=115$, $6$ & $1.00$ & --- & $+0.003$ & $0.02$ \\
$R_X=115$, $6$ & $1.10$ & --- & $-0.010$ & $0.02$ \\
$R_X=115$, $6$ & $1.25$ & --- & $-0.041$ & $0.04$ \\
$R_X=115$, $6$ & $1.50$ & --- & $-1.495$ & $0.17$ \\
$R_X=100$, $6$ & $0.50$ & --- & $+0.128$ & $0.25$ \\
$R_X=100$, $6$ & $0.65$ & --- & $+0.007$ & $0.05$ \\
$R_X=100$, $6$ & $0.80$ & --- & $+0.012$ & $0.01$ \\
$R_X=100$, $6$ & $0.90$ & --- & $+0.002$ & $0.01$ \\
$R_X=100$, $6$ & $1.00$ & --- & $-0.001$ & $0.01$ \\
$R_X=100$, $6$ & $1.10$ & --- & $-0.011$ & $0.02$ \\
$R_X=100$, $6$ & $1.25$ & --- & $-0.264$ & $0.07$ \\
$R_X=100$, $6$ & $1.50$ & --- & $-1.136$ & $0.32$ \\
\end{tabular}
\end{ruledtabular}
\end{table*}

\begin{table}[!ht]
\caption{\label{tab:lr}Light-ring frequencies of the sill ($R_X=100$) and the drain ($A=40$), in units of $10^{-2}$: frequency of the peak response to a drive at $r=110$ (width $4$, amplitude $2\times10^{-4}$, sponge on), on a grid of step $2\%$, averaged over rings at $r=112$--$130$. Only the ratio is used.}
\begin{ruledtabular}
\begin{tabular}{cccc}
$m$ & sill & drain & sill/drain\\
\colrule
3 & $1.56$ & $1.53$ & $1.02$\\
4 & $1.78$ & $1.75$ & $1.02$\\
5 & $2.42$ & $2.37$ & $1.02$\\
6 & $3.05$ & $2.99$ & $1.02$\\
\end{tabular}
\end{ruledtabular}
\end{table}

\begin{table}[!ht]
\caption{\label{tab:bench}Lattice against continuum (Eq.~\eqref{eq:bench} on the measured flows), $m=4$, without the sponge. Ratio: flux next to the horizon per unit depth amplitude at $r=160$, lattice over continuum. $\hat G$: amplitude-normalised gain.}
\begin{ruledtabular}
\begin{tabular}{ccccc}
sill & $f$ & ratio & $\hat G$ measured & $\hat G$ continuum \\ \colrule
$R_X=115$ & $0.5$ & $0.98$ & $+20.2$ & $+19.6$ \\
 & $0.7$ & $0.89$ & $+2.25$ & $+2.47$ \\
 & $0.9$ & $0.89$ & $+0.183$ & $+0.203$ \\
 & $1.3$ & $0.94$ & $-0.138$ & $-0.143$ \\
$R_X=100$ & $0.5$ & $0.79$ & $+7.49$ & $+9.11$ \\
 & $0.7$ & $1.07$ & $+0.662$ & $+0.607$ \\
 & $0.9$ & $0.88$ & $+0.038$ & $+0.042$ \\
 & $1.3$ & $0.94$ & $-0.039$ & $-0.037$ \\
\end{tabular}
\end{ruledtabular}
\end{table}

\begin{table*}[!ht]
\caption{\label{tab:benchnu}Lattice against continuum at three viscosities, $R_X=115$, $m=4$, without the sponge. Ratio: flux next to the horizon per unit depth amplitude, lattice over continuum, normalised at $\rh+5$ and at $r=160$, for the co-rotating wave and its twin. $\hat G(\rh+5)$: ratio of the two fluxes at equal depth amplitude at $\rh+5$.}
\begin{ruledtabular}
\begin{tabular}{cccccccc}
$\nu$ & $f$ & ratio at $\rh+5$, co & twin & $\hat G(\rh+5)$ measured & continuum & ratio at $r=160$, co & twin \\ \colrule
$0.10$ & $0.5$ & $0.98$ & $1.04$ & $1.31$ & $1.40$ & $0.98$ & $0.96$ \\
 & $0.9$ & $2.05$ & $1.01$ & $22.3$ & $10.9$ & $0.89$ & $0.99$ \\
$0.05$ & $0.5$ & $0.91$ & $0.87$ & $1.48$ & $1.42$ & $0.67$ & $0.74$ \\
 & $0.9$ & $1.57$ & $1.00$ & $18.2$ & $11.5$ & $0.85$ & $0.99$ \\
$0.025$ & $0.5$ & $0.67$ & $0.68$ & $1.33$ & $1.35$ & $0.37$ & $0.51$ \\
 & $0.9$ & $1.00$ & $0.98$ & $15.0$ & $14.8$ & $0.78$ & $1.02$ \\
\end{tabular}
\end{ruledtabular}
\end{table*}

\begin{table*}[!ht]
\caption{\label{tab:benchsink}The drain ($A=40$, $r_d=100$, $m=4$) under the checks of Table~\ref{tab:benchnu}. $G$: near-field gain with the sponge off and on. Ratio: flux per unit depth amplitude at $\rh+5$, lattice over continuum, sponge off. Last row: drain of twice the size ($A=80$, $r_d=200$, $N=1536$), sponge on.}
\begin{ruledtabular}
\begin{tabular}{cccccc}
$\nu$ & $f$ & $G$ off & $G$ on & ratio at $\rh+5$, co & twin \\ \colrule
$0.10$ & $0.5$ & $+0.09$ & \revr{$+0.15$} & $0.88$ & $2.40$ \\
 & $0.9$ & $+0.007$ & $+0.004$ & $0.78$ & $1.03$ \\
 & $1.3$ & $-0.14$ & $-0.34$ & & \\
$0.05$ & $0.5$ & $+0.06$ & $+0.16$ & $0.63$ & $3.05$ \\
 & $0.9$ & $-0.001$ & $-0.001$ & $-0.10$ & $1.06$ \\
 & $1.3$ & $-0.14$ & $-0.35$ & & \\
$0.025$ & $0.5$ & $+0.05$ & $+0.16$ & $0.49$ & $2.88$ \\
 & $0.9$ & $-0.011$ & $-0.004$ & $-1.32$ & $1.11$ \\
 & $1.3$ & $-0.14$ & $-0.37$ & & \\
$0.10$, $A=80$ & $0.5$ / $1.3$ & & $+0.14$ / $-0.28$ & & \\
\end{tabular}
\end{ruledtabular}
\end{table*}

\begin{table*}[!ht]
\caption{\label{tab:fj}Single-mode check at the read-out rings: $(F_J/I)\,/\,(m/\omega)$, which is $1$ for a pure $(m,\omega)$ mode. Sills without the sponge; drain $A=40$. At $f=1.0$ for $m=6$ the flux $I$ passes through zero.}
\begin{ruledtabular}
\begin{tabular}{lcccccccccc}
$m=4$, $f$ & $0.3$ & $0.5$ & $0.7$ & $0.9$ & $1.1$ & $1.3$ & & & & \\ \colrule
$R_X=115$ & $0.97$ & $1.07$ & $1.02$ & $1.07$ & --- & $1.00$ & & & & \\
$R_X=100$ & $1.04$ & $1.04$ & $1.07$ & $1.04$ & --- & $0.99$ & & & & \\
drain & $0.92$ & $0.88$ & $0.75$ & $0.97$ & $0.97$ & $0.98$ & & & & \\ \colrule
$m=6$, $f$ & $0.2$ & $0.35$ & $0.5$ & $0.65$ & $0.8$ & $0.9$ & $1.0$ & $1.1$ & $1.25$ & $1.5$ \\ \colrule
$R_X=115$ & $1.02$ & $1.02$ & $1.04$ & $1.01$ & $1.10$ & $1.17$ & $1.38$ & $0.98$ & $0.98$ & $1.00$ \\
$R_X=100$ & $0.98$ & $0.99$ & $0.99$ & $1.11$ & $1.06$ & $1.15$ & \revr{---} & $0.93$ & $0.98$ & $1.00$ \\ \colrule
$m=6$, $f$ & $0.5$ & $0.8$ & $0.95$ & $1.1$ & $1.25$ & & & & & \\ \colrule
drain & $1.18$ & $1.12$ & $1.17$ & $0.99$ & $0.97$ & & & & & \\
\end{tabular}
\end{ruledtabular}
\end{table*}

\end{document}